\documentclass[aps,prl,twocolumn,superscriptaddress,longbibliography]{revtex4-1}

\usepackage[utf8]{inputenc}
\usepackage[english]{babel}
\usepackage[T1]{fontenc}
\usepackage{amsfonts}

\usepackage{amsthm}
\usepackage{amsmath}
\usepackage{bbm}

\usepackage{graphicx}
\usepackage[table]{xcolor}
\usepackage{subcaption}

\usepackage{caption}

\newcommand{\vek}[1]{\mathbf{#1}}

\begin{document}

\title{\textbf{Geometric effects in random assemblies of ellipses}}

\author{Jakov Lovrić}
\affiliation{Division of Physical Chemistry, Ru\dj er Bo\v{s}kovi\'c Institute, Bijeni\v{c}ka cesta 54, 10000 Zagreb, Croatia.}
\affiliation{PULS Group, Department of Physics, Interdisciplinary Center for Nanostructured Films, Friedrich-Alexander-University Erlangen-N\"urnberg, IZNF, Cauerstrasse 3, 91058 Erlangen, Germany.}

\author{Sara Kaliman}
\affiliation{PULS Group, Department of Physics, Interdisciplinary Center for Nanostructured Films, Friedrich-Alexander-University Erlangen-N\"urnberg, IZNF, Cauerstrasse 3, 91058 Erlangen, Germany.}

\author{Wolfram Barfu\ss}
\affiliation{PULS Group, Department of Physics, Interdisciplinary Center for Nanostructured Films, Friedrich-Alexander-University Erlangen-N\"urnberg, IZNF, Cauerstrasse 3, 91058 Erlangen, Germany.}

\author{Gerd E. Schröder Turk}
\affiliation{Murdoch University, College of Science, Health, Engineering and Education, 90 South St, Murdoch WA 6150, Australia}
\affiliation{Niels Bohr Institute and Department of Food Science, University of Copenhagen, Rolighedsvej 26, DK-1958 Frederiksberg, Denmark.}
\affiliation{Division of Physical Chemistry, Lund University, Naturvetarv\"{a}gen 14, SE-221 00 Lund, Sweden.}

\author{Ana-Sunčana Smith}
\affiliation{Division of Physical Chemistry, Ru\dj er Bo\v{s}kovi\'c Institute, Bijeni\v{c}ka cesta 54, 10000 Zagreb, Croatia.}
\affiliation{PULS Group, Department of Physics, Interdisciplinary Center for Nanostructured Films, Friedrich-Alexander-University Erlangen-N\"urnberg, IZNF, Cauerstrasse 3, 91058 Erlangen, Germany.}

\maketitle
\section{Abstract}
Assemblies of anisotropic particles commonly appear in studies of active many-body systems. However, in two dimensions, the geometric ramifications of the finite density of such objects are not entirely understood. To  fully characterize these effects, we perform an in-depth study of random assemblies generated by a slow compression of frictionless elliptical particles. The obtained configurations are then analysed using the Set Voronoi tessellation which takes the particle shape into the account. Not only that we analyse most scalar and vectorial morphological measures, which are commonly discussed in the literature or which have been recently addressed in experiments, but also systematically explore the correlations between them. While in a limited range of parameters similarities with findings in 3D assemblies could be identified, important differences are found when a broad range of aspect ratios and packing fractions are considered. The data discussed in this study should thus provide a unique reference set such that geometric effects and differences from random assemblies could be clearly identified in more complex systems, including ones with soft and active particles that are typically found in biological systems.

\section{Introduction}

Assemblies generated by a finite stochastic process are considered to be random \citep{torquato2000}. They are found in a number of systems including living matter, liquids, granular media, glasses and amorphous solids  \citep{dowsland1992,aste2008a}. 
\begin{figure}[h!]
\includegraphics[width=\linewidth]{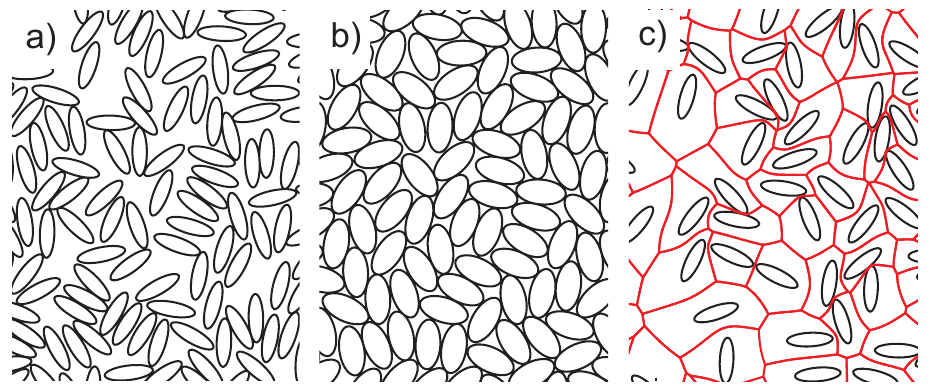}
\caption{Random assemblies of ellipses. a) Ellipses with the elongation $e=3.33$ at a packing fraction of $\phi_g=0.5$ b) Ellipses with $e=2$ at $\phi_g=0.8$ c) Set Voronoi diagram built from the shapes of the generating ellipses with aspect ratio $e=3.33$ at packing fraction $\phi_g=0.2$, clearly demonstrating the non-polygonal nature of of the tessellation.}
\label{fig:fig1}
\end{figure}
In the attempt to reproduce experimentally observed arrangements \cite{scott1960,rutgers1962,farhadi2014,jorjadze2011}, a variety of different algorithms to generate such structures were developed  over the last decades \cite{lubachevsky1990,delaney2005,donev2005a,jia2001,xu2010}, somewhat reflecting the lack of a unique definition of random assembly. It is therefore natural that significant efforts were invested into identifying formally particular states, an example of which is the maximally random jammed state \citep{torquato2000,zong2014}. Furthermore, particular attention was devoted to modeling the state diagrams \citep{donev2007,schreck2010,xingchen2013}, and other properties of random assemblies,  including  correlations between neighboring particles \cite{donev2005b,schaller2015a}, often as a function of the system's dimension, shape of typically hard particles, the assembly packing fraction, and dispersivity \cite{donev2007,farr2009,jin2014,miklius2012,odonovan2013,xingchen2013}.
\begin{figure*}[]
\includegraphics[width=\linewidth]{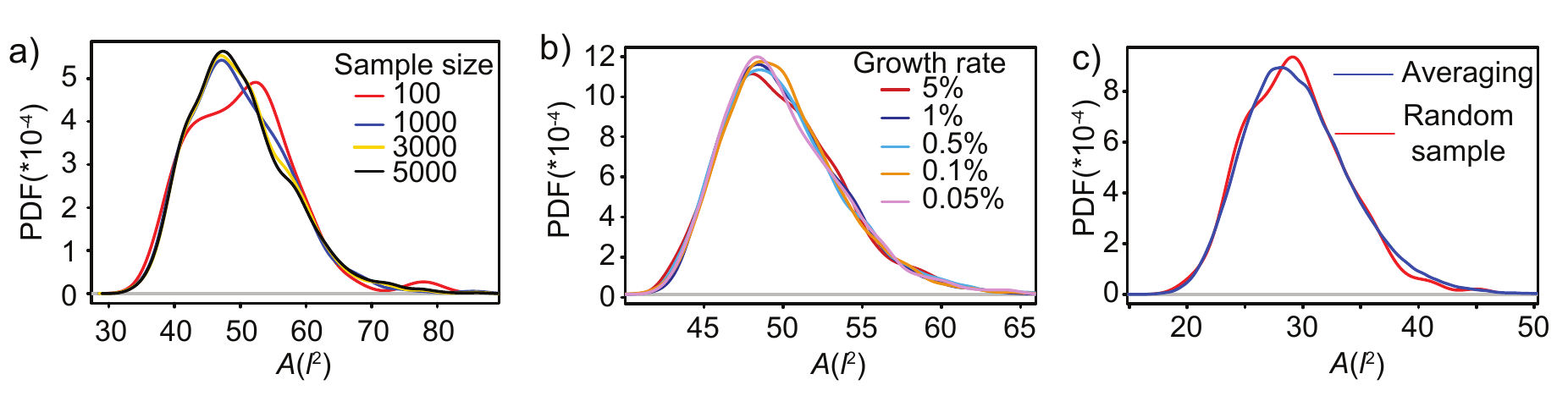}
\caption{Validation of the sampling procedure.  a)  Convergence of the sample size is achieved for sample size $m>1000$. b) Distributions of cell area for packings generated with different growth rate do not statistically differ (characteristic p-value between $0.39$ and $0.96$). c) Distributions of data collected cumulatively (blue line) using all cells from 25 samples of assemblies with 200 ellipses (total of 5000 cells) data) and by sampling of one random cell in from each of 5000 samples of assemblies of 200 ellipses (red dashed line). The comparison of the two approaches shows no statistically significant difference (Kolmogorov-Smirnov test p-value = $0.7$). }
\label{fig:fig2}
\end{figure*}

A common approach to investigate random assemblies is to calculate and study properties of the Voronoi cells generated from the shapes of the particles. In the absence of long-range interactions between particles, it was demonstrated that the so-called Minkowski functionals provide a useful set of structure metrics to characterise the morphology, which are endowed with some completeness with respect to additive properties \citep{arns2004}. Interestingly, for a broad range of systems, a subset of measures consisting of selected Minkowski scalars (area, perimeter, mean curvature) and tensors (moments of inertia) capture the bulk of the information about the assembly \citep{turk2010,turk2013}.  This approach lead to the understanding that in assemblies of circles at high packing fractions (where crystalline order is expected at zero temperature), quenching yields glassy states with a number particles having five and seven neighbours \citep{kumar2005}.  Latter, in hard-sphere systems the number of contacts was calculated as a function of packing fraction \citep{schaller2015a}, and the maximally random jammed packings was fully characterized \cite{klatt2014}. Minkowski functionals were furthermore applied for quantifying structural similarity of crystalline patterns, and used in an accurate identification of the ordering transition and the random close packing limit,\citep{kapfer2012} the later found to be higher than previously determined \citep{berryman1983,donev2004}. 

A number of studies have focused on distributions of various morphological measures characterizing the respective Voronoi cells, the properties of which are imposed by nature of the protocol that produces the assembly. For example, distributions of Voronoi cell volume, area, perimeter and number of neighbours generated by the Poisson \cite{hinde1980,kumar1992} and more ordered point processes in $2$D and $3$D \cite{zhu2001}, were found to be well represented by the Gamma distribution. The cell volumes of monodisperse spheres at low packing fractions also follow the Gamma distribution, while at higher packing fractions the distributions were more narrow and symmetric. The appearance of the Gamma distribution was supported by a theoretical finding that maximizing the entropy of the Voronoi cell volume in random assemblies of spheres in $3$D resulting in cell volume distributions which are a rescaled gamma distribution, the so-called k-gamma distribution \citep{aste2008b}.  

While the analysis of distributions of morphological measures is necessary to understand  of a typical shape of cells in the assembly, it does not provide any information about the organization of these cells\cite{kim2018,kim2019}. Understanding the relationship between the geometry and topology of the assembly actually requires the analysis of correlations between various measures and  the number of neighbours \cite{durand2011,durand2014}.  While second order correlations between the mean of several morphological Voronoi cell measures and the number of neighbours were discovered in tessellations generated from Poisson points \cite{zhu2001}, most works so far focused on first order correlations.  Perhaps the most famous examples of the latter are the Lewis' \cite{lewis1928,lewis1931,chiu1995,kim2015} and the Desh's  \cite{desch1919,rivier1985} laws, predicting a linear increase of the average area and the perimeter of cells with a particular number of neighbours, respectively. Both, of these relationships suggest that large cells have a tendency to have more neighbours, compared to small cells.  Furthermore, the fact that cells with fewer neighbours tend to have neighbours with more sides is captured by the so called Aboav-Weaire's law \citep{aboav1970,aboav1980,weaire1974}. This law was found to change for the diffusion limited colloidal aggregation to an inverse square-root dependence \cite{hilhorst2006,earnshaw1994}. However, the validity of the Lewis, Desh's and Aboav-Weaire's law was confirmed in a number of systems including biological tissues \cite{lewis1928,mombach1990,mombach1993,pina1996,rosa2008,kim2015,kaliman2016}, foam structures \cite{weaire1984,glazier1987,vaz2008}, grain distribution of 2D polycrystalline films \cite{fradkov1985,ciupinski1998}, and even in the analysis of the  polygonal networks on Mars surface \cite{saraiva2009}.  

Here we focus on random assemblies of monodisperse ellipses and systematically scan the whole range of packing fractions and shapes of generating bodies  (Fig. \ref{fig:fig1}a and \ref{fig:fig1}b). We calculate a number of morphological measures of Voronoi cells, including the most usually studied area $A$, perimeter $P$, and the number of neighbours ($n$). Motivated by the recent work on phase transitions in polygonal networks \cite{bi2015}, the mean area of all assemblies is kept constant. Moreover, inspired by the previous work on Minkowsky functionals \citep{turk2010,turk2013}, we use elongation $E$ to capture anisotropic effects, along with the standard deviation in the contact length ($SCL$). To explore the centrality of the generated assemblies, we investigate the distance of between the center of mass of the generating ellipse and the center of mass of the Set Voronoi cell $CMD$ \cite{klatt2019}.  Last but not least, we introduce and investigate the behaviour of rescaled measures in which all lengths are rescaled by the square root of the area of the respective cell. The distributions and correlations between measures are calculated as a function of packing fraction and the elongation of the generating ellipse, which allows us to put the results obtained for assembled ellipses in the context of hard spheres and Poission points. By this we provide a comprehensive set of characteristics for measures that are available by direct imaging methods, and have been investigated in a broad range of problems.
\begin{figure*}[]
\includegraphics[width=0.85\linewidth]{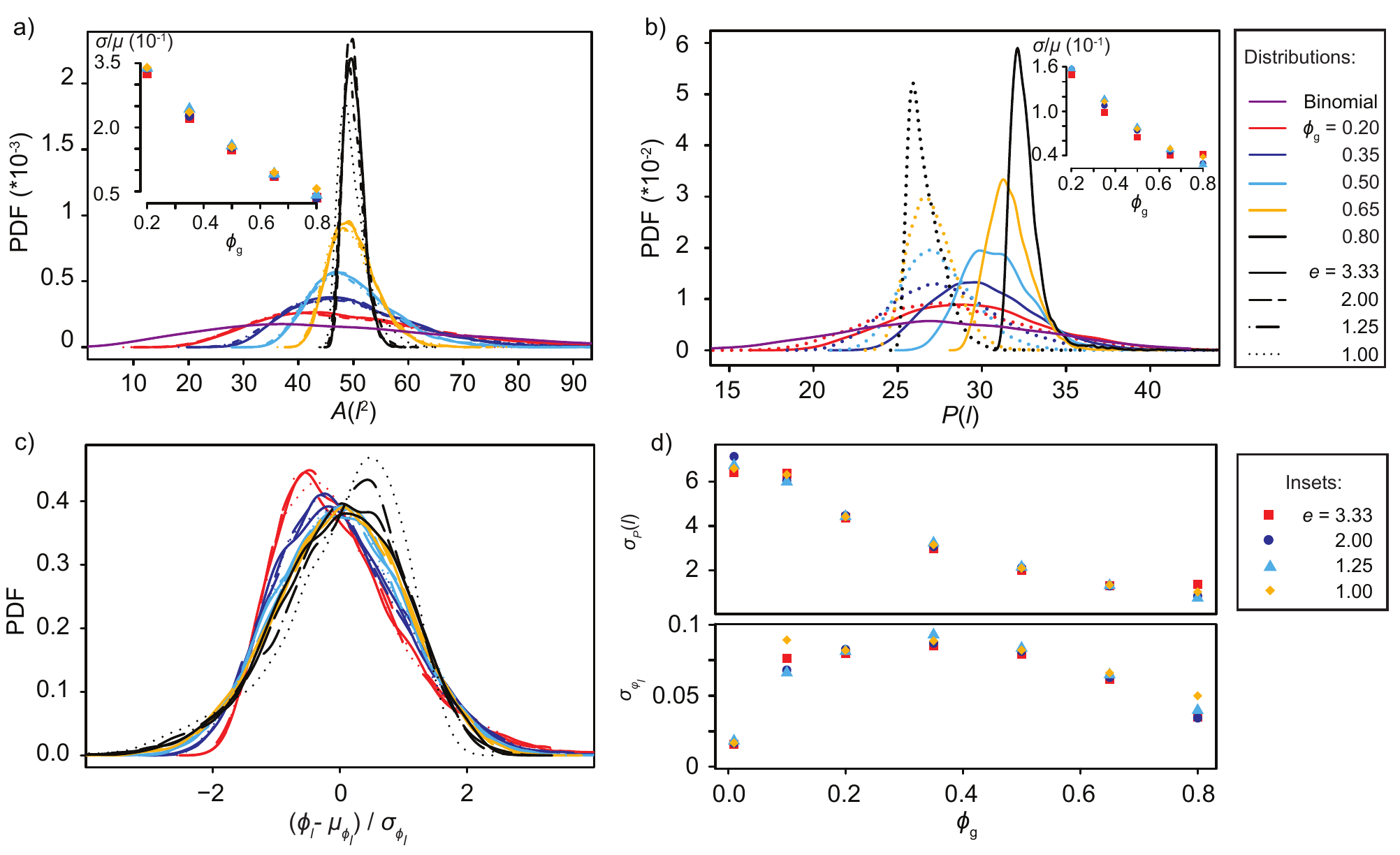}
\centering
\caption{Characteristics of Voronoi cell area and perimeter. a) Cell area distributions. Insert shows coefficients of variation of cell area distributions. b) Cell perimeter distributions. Insert shows coefficients of variation of cell perimeter distributions. c) Standardized distributions of the local packing fraction. Distributions overlap at packing fractions $\phi_g = 0.5, 0.65$ with $ p \text{-value}\in [0.02, 0.97]$ d) Standard deviation of the distribution of the local packing fraction and  cell perimeters are shown in the upper and lower panel, respectively, as a function of the global packing fraction, for different shapes of ellipses.}
\label{fig:fig3}
\end{figure*}
\section{Methods}

\subsection{Parameter space and image analysis}
We characterize the shape of the particle by its aspect ratio $e$, the latter being defined as the ratio between the length of the longer and the shorter axis of the ellipse. Accordingly, $e$ is greater than $1$ and adopts large values for elongated bodies and is equal to unity for circles.  We use monodisperse particles with  $e = 3.33, 2, 1.25, 1$ to generate assemblies with global packing fractions $\phi_g = 0.2, 0.35, 0.5, 0.65, 0.8$ before jamming (typical examples shown in Fig. \ref{fig:fig1}a, \ref{fig:fig1}b and ESI,\dag Figs. 1-5). When computing Voronoi tessellations we use the so-called Set Voronoi tessellation \citep{schaller2013,weis2017}, also known as navigational map or as tessellation by zone of influence. This definition determines the Voronoi tessellation taking particle asphericity and shape into account. Cells are defined by proximity to the particle bounding surface, rather than the particle center.  The Voroni cell $V(E_i)$ of particle $E_i$ is the region of space defined as follows:
\begin{equation}
V(E_i) =\{r\in \mathbb{R}^2|d(r,E_i)\le d(r,E_j), \forall j\ne i, j\in\{1,\ldots,n\}\}
\end{equation}
where $d$ is the Euclidean metric. Voronoi cell of $i$-th particle $V(E_i)$ contains all points $l$ in plane that are closer to $i$-th particle than to any other particle $j$ from the set of $n$ particles. The result of this tessellation are Voronoi cells  with edges that can adopt arbitrary curvatures (Fig. \ref{fig:fig1}c). 

All morphological measures are extracted from output of simulations using the procedure described in \citep{schaller2013}.
 \begin{figure*}[]
\includegraphics[width=.75\linewidth]{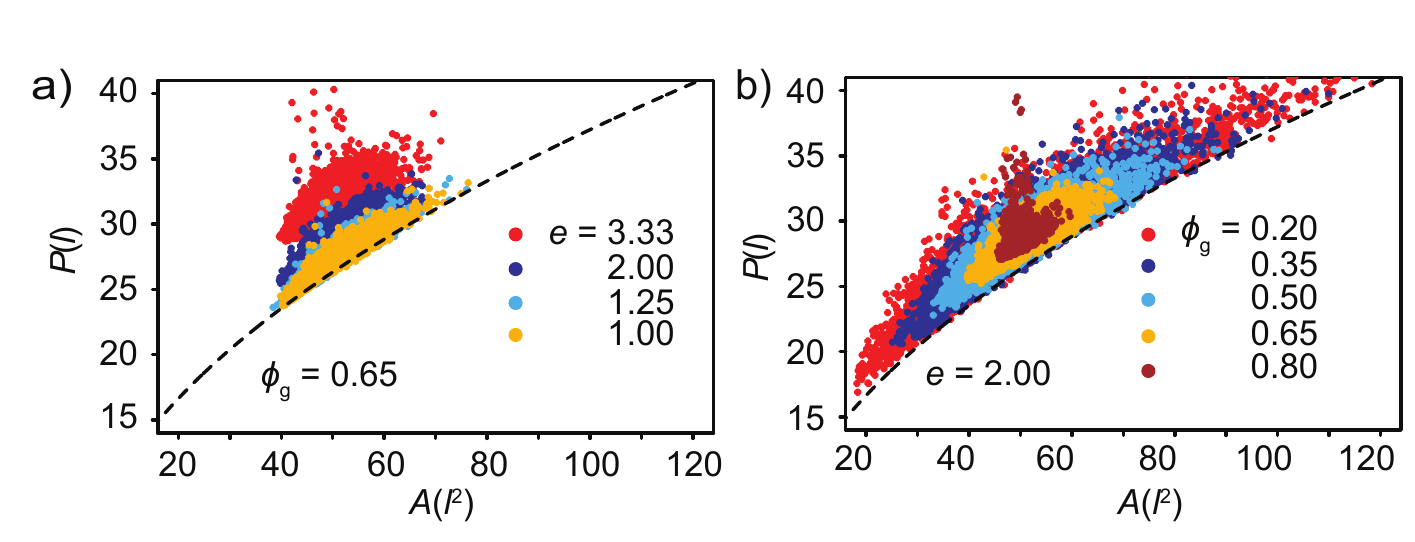}
\centering
\caption{Correlations between cell area and perimeter. a) Scatter points displaying cell perimeters as a function of their area are shown  for fixed packing fraction $\phi_g = 0.65$, and  b) for a fixed aspect ratio $e = 2$. Each cell is represented by one point.  The line represent the dependence of the perimeter of a hexagon as a function of its area $P_{hex}=(8\sqrt{3}A_{hex})^{1/2}$, calculated with the same resolution as the Voronoi tessellations. c) Map of the Pearsons correlation coefficient  for the available global packing fractions and and shapes of packed particles.}
\label{fig:fig4}
\end{figure*}

\subsection{Generation of assemblies and sampling procedure}

To generate random assemblies we use a slightly modified algorithm originally developed by Delaney \textit{et al}. \cite{delaney2005}. Initially, a chosen number of ellipses (200) are reduced to $20 \%$ of their actual area and are randomly placed into the simulation box. If there is overlap between ellipses, one ellipse is randomly chosen (equal probability for each member of the set), and a random translation and rotation is performed. The translation  is defined by a vector whose length is drawn from the uniform distribution on the interval $[-\frac{b}{2},\frac{b}{2}]$ where $b$ is semi-minor axis of the ellipses at full size. Translations in $x$ and $y$ directions are executed independently. The value of the rotation angle is a random variable, drawn from  is drawn from the uniform distribution on the interval $[0,\pi]$. The movement (2 translations + rotation) is accepted if the total overlap decreases or stays the same. The procedure is repeated until the total overlap is  $0$ (minimal significant value in double floating precision arithmetics), when all ellipses are simultaneously dilated by $0.5\%$ of their final size. If the total overlap caused by the growth is greater than $0$, then translations and rotations of ellipses are repeated until the overlap vanishes again.  

All simulations are performed with periodic boundary conditions and with the simulation box of the predefined resolution. This ensures that the mean area of Voronoi cells is constant for all data and, hence, that morphological measures are all calculated with the same accuracy, irrespective of the choice of parameters $e$ and $\phi_g$. 

Convergence of the sample is achieved after generating 25 assemblies containing 200 ellipses  (Fig. \ref{fig:fig2}c), which provides distributions with $5000$ entries. We find that the  assemblies are by and large insensitive to the growth rate of the ellipses (Fig. \ref{fig:fig2}b) as well as to details of distributions from which translations and rotations are generated. Notably, this is different to the Lubachevsky-Stillinger algorithm where the growth rate tunes the final packing density. Last but not least, we verify that it is identical to draw 5000 times one randomly chosen cell from 5000 realizations of the system containing 200 ellipses (so-called random sampling) as it is to use all cells in the 25 assemblies to achieve the same sample size (Fig. \ref{fig:fig2}a).

\subsection{Statistical analysis}

From each tessellation, for a number of continuous measures $K$ (such as the area, perimeter and others), we construct a sample $\vek{k}=\{k_i, i\in\{1,\ldots, m\}, k_i\in\mathbb{R}\}$, where  $\vek{k}$ is a 1D array containing information on the magnitude $k_i$ of the measure $K$ for each cell denoted by the index $i$. Each measure is then probed with the generalized gamma density function  $f(k_i|\theta)$

\begin{equation}
f(k_i|\theta)=\frac{\tau}{\lambda\Gamma(\alpha)}\Big(\frac{k_i-k_0}{\lambda}\Big)^{\alpha\tau-1}\text{e}^{-(\frac{k_i-k_0}{\lambda})^{\tau}}\mathbbm{1}_{[k_0,+\infty>}(k_i), 
\label{eq:gengamma}
\end{equation}
which is parametrized by  $\theta=(\alpha$, $\tau$, $\lambda) \in \mathbb{R}_{+}^{3}$ , and $k_0\in \mathbb{R}$. Here, $\Gamma(.)$ denotes a Gamma function.  With $\tau =1$ one recovers the usual Gamma distribution, $\alpha =1$ give the Weibull distribution, while $\alpha=\tau =1$ provides an exponential distribution. Applying eq. 2 to distributions obtained in simulations, therefore, allows us to identify the distribution from a fit. For the number of neighbors $m$, which is a discrete variable, the distribution is calculated following $\mathbbm{P}(n)=\mathbbm{P}(Y\in \langle n-0.5,n+0.5]), n \in \mathbb{N}$, where $Y$ is a random variable with the continuous distribution. 

In all cases, the fit relies on the maximum likelihood estimator (MLE), denoted by $\hat\theta$
\begin{equation}
\hat\theta =\max\limits_{\theta\in \mathbb{R}_{+}^{3}}L(\theta) \text{ where } L(\theta) =\prod\limits_{i=1}^mf(k_i|\theta).
\end{equation}
The MLE can be considered as an estimator which maximizes the probability of obtaining sample $\vek{k}$ over the parametric space $\mathbb{R}_{+}^{3}$. The goodness of the fit is tested with Pearsons-$\chi ^2$ test (least squared method relative to estimated density function) while equality of distributions is tested by Kolmogorov-Smirnov test, both at confidence level of $0.01$.

By denoting the mean of the set as
\begin{equation}
\bar{k} = m^{-1}\sum\limits_{i=1}^mk_i,
\label{eq:corr}
\end{equation}
it is possible to determine the so-called Pearsons coefficient $\rho$ which quantifies correlations between morphological measures $K$ and $H$ represented by samples  $\vek{k}$ and $\vek{h}$ 
\begin{equation}
\rho = \frac{\sum\limits_{i=1}^m(k_i-\bar{k})(h_i-\bar{h})}{\sqrt{\sum\limits_{i=1}^m(k_i-\bar{k})^2}\sqrt{\sum\limits_{i=1}^m(h_i-\bar{h})^2}}
\end{equation}
This is particularly meaningful when the relation between two measures is linear.

\begin{figure*}[]
\includegraphics[width=0.77\linewidth]{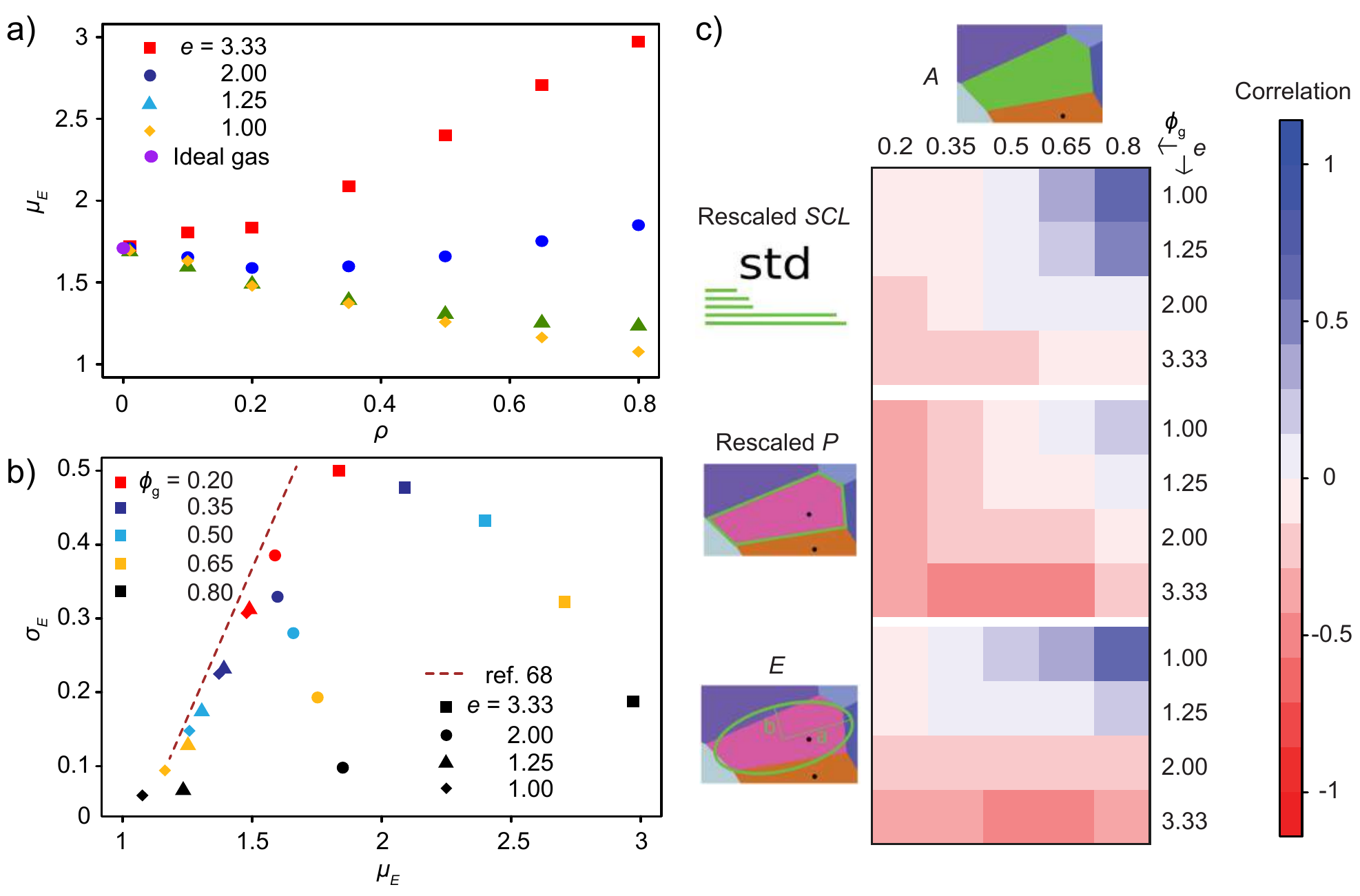}
\centering
\caption{Average elongation and its correlations with other morphological measures. a) Mean elongations for all of considered assemblies. b) Relation between mean and standard deviation of elongation for all of the considered assemblies. Each global packing fraction is associated with a particular color, while the symbol shape signifies a particular shape. c) Correlations between the area and rescaled versions of morphological measures of shape. All of the morphological measures of shape have similar correlations with the cell area.}
\label{fig:fig5}
\end{figure*}

\section{Results}

\subsection{Distributions and correlations between Voronoi area and perimeter}

We first analyse distributions  characterized by their mean $\mu$ and standard deviation $\sigma$, of the areas and perimeters of Voronoi cells for all generated assemblies (Fig. \ref{fig:fig3}). We find that generalized gamma distribution (Equation \ref{eq:gengamma}) fits Voronoi cell area nicely with parameter $\tau \approx 1$ for most packings at packing fractions below $0.8$. Therefore, we set parameter $\tau=1$ in our fits and in agreement with previous work on cell area distributions emerging from a Poisson process \cite{hinde1980,kumar1992,lazar2013,aste2008b}, and ellipsoidal assemblies in 3D \cite{schaller2015b,schaller2016}, find gamma distributions in the bulk of the phase space ($\phi_g<0.8$ and all $e$). In this regime the $\chi^2$ test typically gives $p$-values significantly larger than 0.1 and deviations between the fit and the sampled distributions have no structure (see ESI,\dag Figs. 6-10 for details). Importantly, this level of significance is obtained only if all parameters of the Gamma distribution fit ($\alpha$,$\lambda$,$x_0$), are left free, allowing $\alpha$ not adopt integer values suggested previously \cite{aste2008b}. Notably, however, $99\%$ of the data fall within one order in cell area  magnitude.  

Significant deviations from the Gamma distributions are, nonetheless, found for disks at packing fractions close to the jamming transitions (e.g $\phi_g=0.8$ in Fig. \ref{fig:fig3} and SI-Fig.10c,d ), the latter expected at $\phi_g=0.82 \pm0.02$  \citep{berryman1983}. In this regime, the nature of the emergent distributions could not be determined with statistical accuracy, presumably because of the appearance of locally ordered domains in the assembly (blue arrow in ESI,\dag Fig. 5d).

While by construction all distributions have the same mean, the increase of the global packing fraction results in more narrow distributions (Fig. \ref{fig:fig3}a), with the shape of packed ellipses having no significant effect below $\phi_g=0.8$. This is also evident from the inspection of the coefficient of variation ($\sigma/\mu$), shown in the inset of Fig. \ref{fig:fig3}a,  which linearly decreases with $\phi_g$, in a shape-independent manner.

Similarly to cell area, and as found for the Poisson process \cite{hinde1980,kumar1992}, the distribution of Voronoi cell perimeters are also characterized by the Gamma distribution (see ESI,\dag Figs. 11-15 for details). However, in this case, the shape of  the ellipses strongly affects the mean and the width $\sigma$ of the distribution already at global packing fractions above   $\phi_g=0.2$. (Fig. \ref{fig:fig3}b). Besides the narrowing of the distributions with the increased global packing fraction, we find that more elongated ellipses tend to have cells with a greater perimeter on average. That can be explained by the fact that at higher packing fractions, the shapes of the cells gradually start to adopt the shape of the generating ellipse.  Nonetheless, the coefficient of variation still manifests a shape-insensitive decay with the increase in the global packing fraction (inset in Fig. \ref{fig:fig3}b)).

While the distributions of area and perimeters provide a notion of general geometrical properties of the assembly, the variability in the local structure is characterized by  the distribution of local packing fraction ($\phi_l$). The latter is defined as a ratio of particle area $A_{E}$  and corresponding Voronoi cell area $A_{V}$  \cite{schaller2015b}
\begin{equation}
\phi_l=\frac{A_{E}}{A_{V}},
\end{equation}
and reflects the local density of particles in the assembly.

For 3D assemblies of oblate ellipses, normal distribution was found as a good fitting model for the standardized local packing fraction, the latter being  is obtained by first subtracting the distribution mean from each observation and then by dividing the result by the standard deviation of the distribution. Remarkably, such distributions were found to be invariant to packing fraction and shape of the packed particles at packing fractions $\phi_g \in [0.55, 0.72]$ in 3D \cite{schaller2015a}, where jammed disordered configurations form. In the current  assemblies, similar behaviour, yielding the overlap of standardized distributions of the local packing fraction is observed only for intermediate global packing fractions (see $\phi_g=0.5$ and $0.65$ in Fig.\ref{fig:fig3}c). However, because local order promotes cells with high local packing fractions (blue arrow in ESI,\dag Fig 5d),  standardized distributions become negatively skewed  at high $\phi_g$ (exemplified for $\phi_g=0.8$ in Fig. \ref{fig:fig3}c). Naturally, these deviations from normality are more pronounced for less elongated objects and disks.
 \begin{figure*}[]
\includegraphics[width=\linewidth]{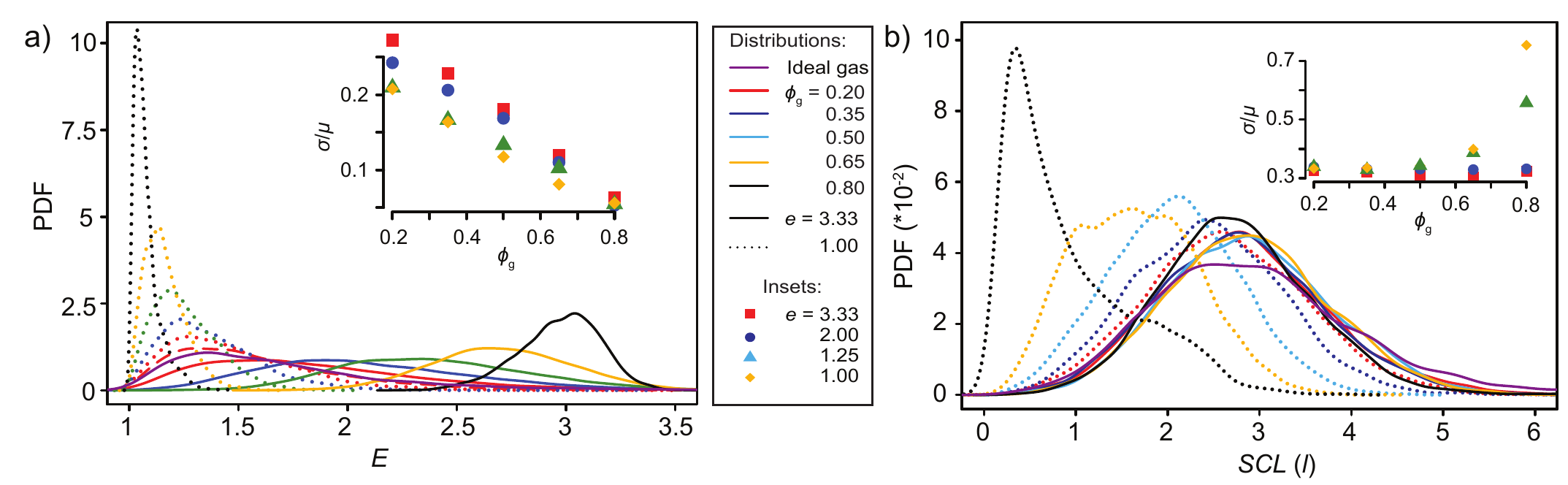}
\caption{Distributions of measures associated with cell anisotropy. In the legend, colour indicates different packing fractions, solid lines are data for $e=3.333$ data and dotted lines for $e=1.00$. a) Distributions of elongation. Insert shows coefficients of variation of cell elongation distributions.  b) Distributions of the standard deviation of contact length. Insert shows coefficients of variation of standard deviation of contact length distributions.}
\label{fig:fig6}
\end{figure*}

At the lowest global packing fractions ($\phi_g=0.2$ in Fig. \ref{fig:fig3}c), the standardized distributions of the local packing fraction are positively skewed meaning there are more cells with a local packing fraction smaller than the average. Then, because of similarity between low-density elliptical assemblies and the Poisson point process, Voronoi cell areas of  ellipses which are close to each other are greater than the average, while their local packing fraction is lower (blue arrow in ESI,\dag Fig 1a). 

The standard deviation of the distribution of local packing fractions $\sigma(\phi_l)$,  due to homogeneity of standard deviation ($\sigma (cK) = |c|\sigma (K), \forall c\in\mathbb{R}$) and the fact that our system is monodisperse becomes
\begin{equation}
\sigma(\phi_l)=\sigma\Big(\frac{A_{E}}{A_{V}}\Big)=A_{E}\sigma\Big(\frac{1}{A_{V}}\Big).
\label{eq:sdlocarea}
\end{equation}
At small  $\phi_g$, the assembly adopts a structure similar to that generated by a Poisson point process for which $A_E=0$. Hence,   as $\phi_g\rightarrow 0$, $\sigma(\phi_l) \rightarrow 0$. On the other hand, at high $\phi_g$, as the  maximum packing fraction is  approached where   $\sigma\big(\frac{1}{A_{V}}\big) \rightarrow 0$,  $\sigma(\phi_l)$ decays nearly linearly (Fig. \ref{fig:fig3}d).

A maximum in the standard deviation of local packing fraction appears at at $\phi_g\simeq0.35$, at  the crossover between trends associated with these two limits. This non-monotonous behaviour is contrasted by previously reported continuous linear decay of $\sigma(\phi_l)$ as a function of  $\phi_g$ for assemblies of 3D oblate ellipsoids \cite{schaller2015a}, although that study considered only  mechanically jammed ellipsoid configurations, with a much smaller range of packing fractions.

Unlike the distributions of the local packing fractions,  the standardized distributions  of the perimeter are very similar for all assemblies  (ESI,\dag Fig 16b), hence a monotonically decreasing standard deviation of the true distribution (Fig. \ref{fig:fig3}d). The exception is the significantly more positively skewed distribution for assemblies at high packing fraction (e.g. $\phi_g=0.8$).  This suggests that a large number of cells associated with ellipse assemblies at high packing fraction have perimeters smaller than the average one, an effect that is most likely a result of counterbalancing the appearance of a small number of structural defects (holes), in a vicinity of which a small number of cells have areas larger than the average (brown arrow in ESI,\dag Fig 5d). 
\begin{figure*}[]
\includegraphics[width=\linewidth]{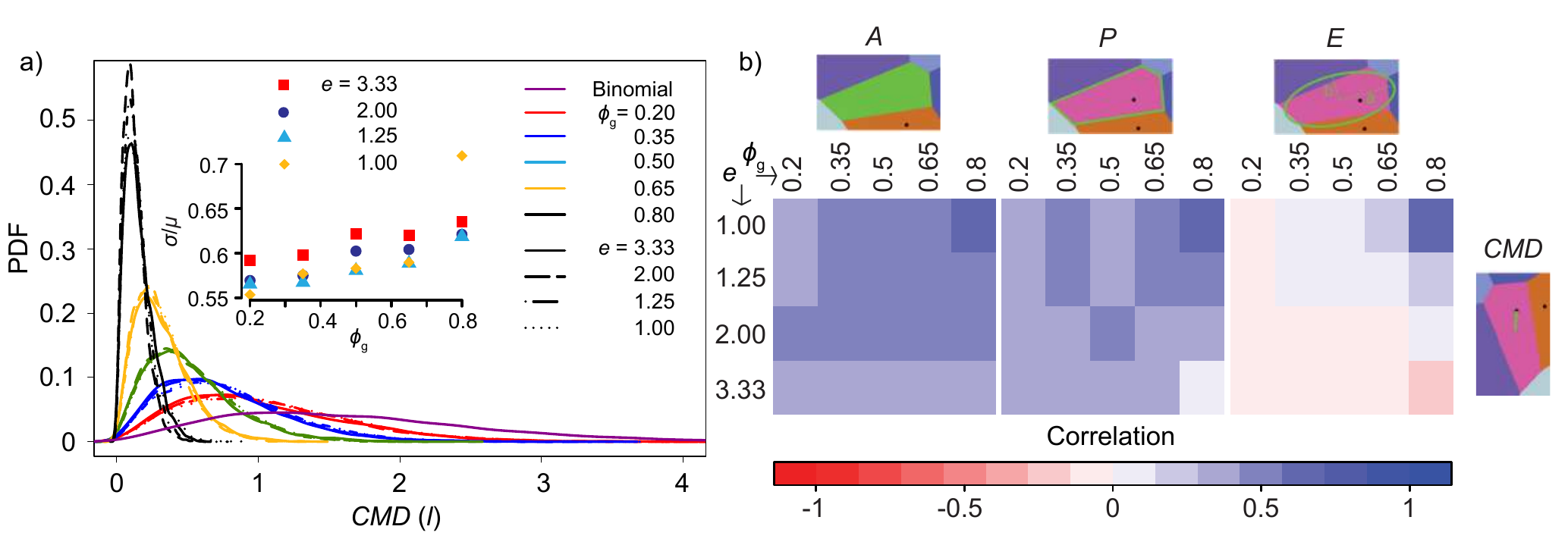}
\caption{Distance between the center of the generating ellipse and the center of the Voronoi cell ($CMD$) as a morphological measure of centrality. a) Distributions of $CMD$ as a function of the packing fraction (indicated by colours, as shown in the legend), and elongation of ellipses (indicated by the curve type). Insert shows $CMD$ coefficients of variation as a function of the global packing fraction (the generating ellipse elongation is denoted by a symbol of different shape and colour). b)Correlations between $CMD$ and cell area, perimeter and elongation.}
\label{fig:fig7}
\end{figure*}

Besides studying individual properties of parameters characterizing cells from the Voronoi construction, it is important to understand the relation between these parameters. For example, a square root relation between cell area and perimeter can be anticipated a priori.  Indeed, such dependence can be inferred from Fig.\ref{fig:fig4}a,b where all data points follow the line showing the dependence of the hexagon perimeter on its area.  Naturally, clouds associated with assemblies of more elongated objects have their center further away from this line. From the shape of clusters, it is, however, evident that both the global packing fraction and the aspect ratio of packed particles have a substantial impact on the shape of the cell characterized by the pair of measures. Specifically, the width (although not the length)of the clusters increases with increasing the elongation of generating particles, suggesting that there is larger diversity of Voronoi cell shapes that can be found in a sample built from more elongated objects (Fig. \ref{fig:fig4}a).  On the other hand, increasing the global packing fraction (Fig. \ref{fig:fig4}b) results in less dispersed clusters, yet the centres of the cluster remain basically at the same position. This suggest that phase space of the accessible shapes simply continuously decreases as the density of ellipses increases. This result could not be anticipated from the analysis of the distributions of cell perimeters or areas alone.

Not surprisingly, strong positive correlation between cell area and perimeter has been determined for all assemblies, simply because larger cells tend to have greater perimeter. However, the correlations between $A$ and $P$ decrease with increasing the global packing fraction and the elongation of the particles.  These results are contrasted by the correlations between the cell area and the so called rescaled perimeters (square root of the isoperimetric ratio for each cell) (Fig. \ref{fig:fig5}c). The later are dimensionless numbers calculated by dividing the original perimeter of the cell with the  square root of the its area. In such a representation, the lower bound of the rescaled perimeter is 3.54 which is the rescaled perimeter of a circular object, while a hexagonal object would be have a rescaled perimeter of 3.72.  Interestingly, in most of the parameter space, we find weak negative correlations between the rescaled  perimeter and the cell area, suggesting that larger cells are more hexagonal than smaller cells. This trend is clearly violated  for disks packed at high global packing fractions  (top right corner of the correlation matrix), where ordered domains, in which cells with high local packing factions (cells  smaller than average) adopt locally hexagonal structures (blue arrow in ESI,\dag Fig 5d), occure.

\subsection{Morphological measures of anisotropy}

The scalar nature of the cell perimeters and area does not allow for the direct inference of the anisotropy of the cells emerging from the tessellation. Nonetheless, cast into the rescaled perimeter, they may provide useful insight. However, the most natural measure of anisotropy is the cell elongation, which is a ratio of the two principle moments of inertia calculated under the assumption of the uniform distribution of mass over the area.  Furthermore, the analysis of the tessellations in epithelial  tissues  identified the standard deviation of contact lengths (SCL) as a measure that reflects the anisotropy of the tessellation \cite{kaliman2016}. Namely, because the contact length is  a section of the perimeter shared by two neighbouring cells, and because the mean number of neighbours is approximately constant across the whole range of packing fractions, more elongated Voronoi cells have both short and long edges which yields the greater standard deviation of contact lengths.

Further evidence that these three morphological measures all capture the anisotropy of Voronoi cells emerges from the analysis of their correlations with the area (Fig. \ref{fig:fig5}c). At low densities and high elongations of generating ellipses, the area and the elongation of Voronoi cells are negatively correlated, while the opposite is true the closer the shape of generating bodies become more disks-like and the higher the concentrations.

In previous study on 3D ellipsoid assemblies, Schaller \textit{et al} found a complex dependence of the mean elongation of the Voronoi cells on the global packing fractions \cite{schaller2015b}. Our results in 2D show similar trends (Fig.\ref{fig:fig5}a). As the maximum packing fractions is approached for disks, the Voronoi cells adopt more isotropic hexagonal shapes resulting in a continuous decrease of mean elongation. The opposite trend is found for highly elliptical objects, suggesting that a t higher densities the shape of the cell and the generating body become  more alike.  Actually, as $\phi_g \rightarrow 0$, mean cell elongations approach value for Poisson point process ($\approx 1.71$ in 2D), and as $\phi_g \rightarrow 1$, mean cell elongation of Voronoi cells approaches the aspect ratio $e$ of packed particles (Fig.\ref{fig:fig5}b). 

It is also interesting to discuss the relation between the mean elongation and standard deviation of elongation, since recently, their linear interdependence was discovered in epithelial tissues \cite{atia2018}. However, it was not clear if this dependence is a result of an active regulation or merely a consequence of assembly of cell nuclei in a plane. In our random assemblies, we find that the standard deviation of elongation  systematically decreases with the global packing fraction (at constant shape of the packed objects). As the shapes approach  $e\simeq1.71$ from above, the standard deviation decreases with the mean elongation. However, once more circular shapes are used to generate assemblies ($e<1.71$), the standard deviation increases as a function of mean elongation (assemblies are more Poisson like), and a linear dependence is recovered, as a property of the average shape of the packed objects.  

Besides the first and the second moment, we also analyse the functional form of true and standardized distributions of measures of anisotropy (Fig. \ref{fig:fig6} and ESI,\dag Fig. 16). Naturally, with the increase of the global packing fraction all distributions become narrower and the aspect ratio of the packed ellipses starts to influence the distributions, as discussed in the analysis of the mean and the variance (Fig. \ref{fig:fig5}). Standardized distributions of elongation have the same features as the distributions of rescaled perimeters, which is further corroborating the relation between these measures. Interestingly however, the standardized distribution of the average contact lengths all become Gaussian-like, except for disks at high densities (ESI,\dag Fig 16b,16d,16f).

\subsection{Measuring cell centrality}

Distance between centers of mass of Voronoi cells (CM-distance) and their generating ellipses reflects the centrality in the assembly \cite{burns2009,du1999, klatt2019}. In fully disordered systems, example of which is the Poisson point process, the distributions of CM-distance are wide (purple line in  Fig. \ref{fig:fig7}a). In most ordered systems, such a crystals, the centers of objects and the centers of Voronoi cells coincide. In this case the distribution of CM-distance distribution is a delta function, and the tessellations are called centroidal Voronoi diagrams. In random assembly of ellipses, the global packing fraction seems to be the main determinant of the  of centrality of the assembly, evident by the narrowing down of the distributions and the shift of the mean towards the $0$ value. From the correlation matrices (Fig. \ref{fig:fig7}b), furthermore,  we observe positive correlations of the CM-distance with  cell's area and perimeter, and no correlation with the elongation, except for circles at high packing fraction which can also be attributed to defects in regular hexagonal assembly (Fig 5d brown arrow).

\subsection{Neighboring and the relation between geometry and topology in assemblies of ellipses}
\begin{figure*}[t]
\centering
\includegraphics[width=0.8\textwidth]{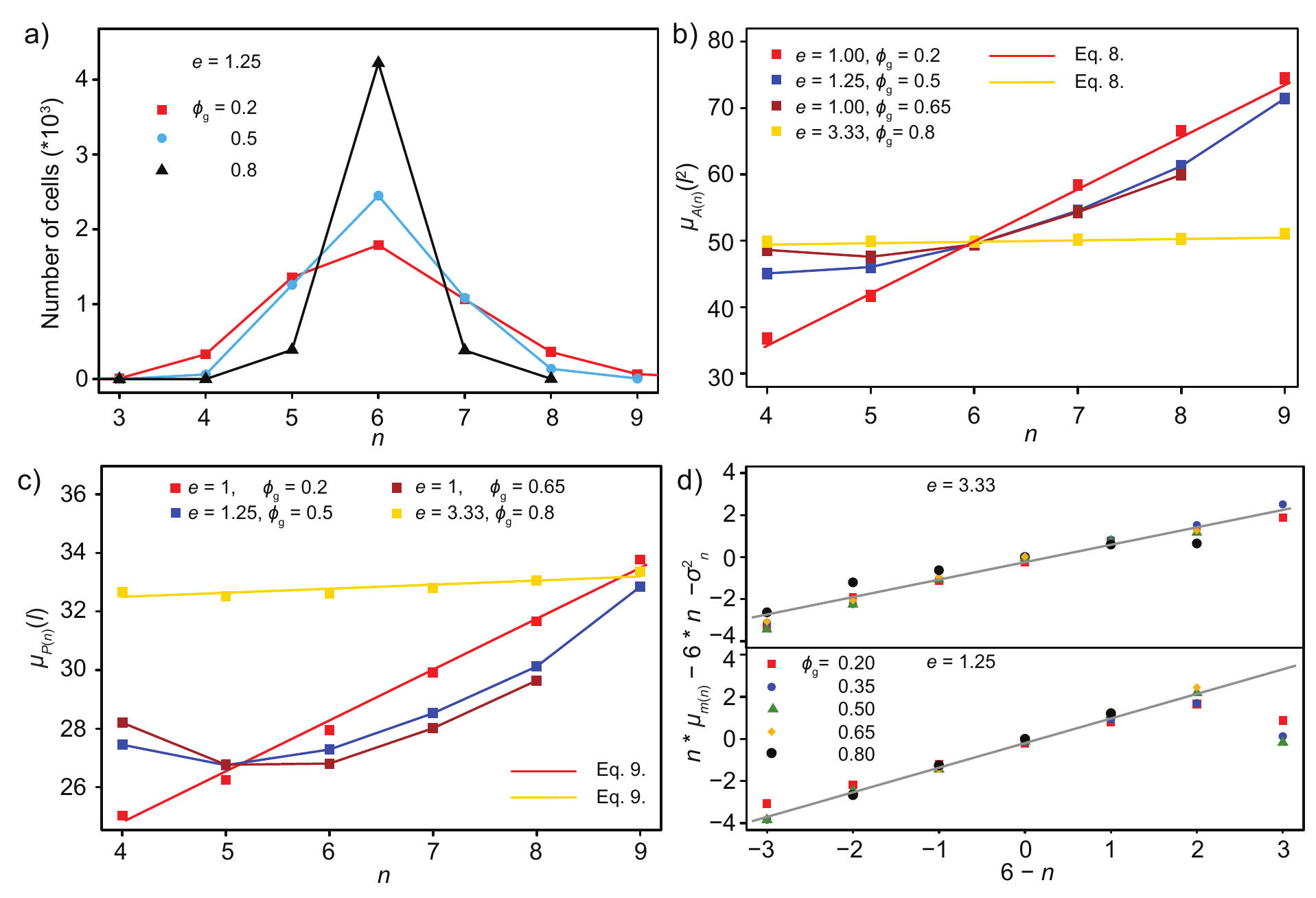}
\caption{ Characterization of number of neighbours as a morphological measure and its correlations. a) Histograms of number of neighbours for ellipses of aspect ratio $e = 1.25$. Different colours and symbols represent different global packing fractions (symbols are data points, while lines are guidance for the eye). The total sample size is $5000$. b) Evaluation of the Lewis' law in different parts of the parameter space spanned by $e$ and $\phi_g$. Symbols denote data points. Red and yellow curves are  linear fits to data using eq. 8. Brown and blue lines are guidance for the eye. c) Evaluation of Desh's law. The demarcation of lines is identical to the graph in b), except for the fit that is here performed using eq. 9. d) Evaluation of the Aboav-Weaire law for ellipses with $e = 3.33$ (top graph) and $e = 1.25$ (bottom graph). Different symbols denote different packing fractions, while the gray line represent a fit using eq. 10.}
\label{fig:fig8}
\end{figure*}
The last measure that we analyse is the number of neighbours of  cells  in  assembly. The extracted distributions  are well represented by the discrete version of the gamma distribution except at high packing fractions were the contribution of cells with less than five and more than seven neighbours is negligible. This behaviour emerges gradually with an increase of the global packing fraction, due to the narrowing of otherwise systematically asymmetric distributions  (Fig. \ref{fig:fig8}a). Actually, the number of cells with $5$ neighborus is  statistically significantly  larger than the number of cells with $7$ neighborus. Nonetheless, the difference is small, which may the the cause of the disagreement with previous reports based on smaller sampling.

We furthermore explore correlations with the number of neighbours, which give insights on the relation between the geometry and topology of the assembly. The most studied example of such a correlations is the Lewis' law \cite{lewis1928,lewis1931,chiu1995,kim2014} which describes a linear relationship between the number of neighbours  $n$, and the average area $\mu_{A(n)}$ of a cell with $n$ neighbours following
\begin{equation}
\mu_{A(n)}=\mu_A\left [1+\alpha(n-6)\right].
\end{equation}
 Here $\mu_A$ is the average area of the cell in the assembly and $\alpha$ is a constant.
 
 A similar expression 
 \begin{equation}
 \mu_{P(n)}=\mu_P\left [1+\beta(n-6)\right], 
 \end{equation}
 exists for the average perimeter $\mu_{P(n)}$ of cells with $n$ neighbours and is know as the Desh's law \cite{desch1919,rivier1985}. Both, of these laws suggest that large, and inversely, small cells have a tendency to have more or less neighbours, respectively. 

In assemblies of ellipses  we observe that Lewis' and Desh's laws  (Fig. \ref{fig:fig8}b,c) have a same range of validity. This is of course promoted by the strong positive correlations between the cell area and cell perimeter. Specifically, at low packing fractions and independent of particle shape, strong linear dependence  (e.g. $e=1$ and $\phi_g = 0.2$ in Fig. \ref{fig:fig8}b) can be confirmed in agreement with previous reports \cite{lewis1928,lewis1931}. Indeed, it is possible to observe in those assemblies that cells with more neighbours have on average greater cell area and perimeter (ESI,\dag Fig 1b,c). However, deviations from the Lewis' and Desh's laws is confirmed at intermediate packing fractions ($\phi_g = 0.5, 0.65$), where the data are more consistent with a quadratic, rather than a linear relation (Fig. \ref{fig:fig8}b,c). This second order dependence was also predicted by calculating the probabilities of establishing certain size and certain number of neighbours \cite{kim2014}. 

Interestingly, however, the average area and the perimeter of the cell becomes, by and large, independent of the neighbourhood for highly elongated particles at high packing fractions (e.g. $e=3.33$ and $\phi_g = 0.8$ in Fig. \ref{fig:fig8}b,c). Consequently,  Lewis' and Desh's law are recovered but with no statistically significant slope. This effect is exemplified in SI. Fig. 5a showing an assembly containing a cell with 8 and 4 neighbours, both having nearly equal areas and perimeters. It is important to notice that this result is a consequence of the set Voronoi tessellation. Actually, liner dependence with a strong slope is recovered for the center-of-mass based tessellation performed on he same data set (ESI,\dag Fig. 17), as suggested previously \cite{lewis1928,kim2014}.

The fact that cells with fewer neighbours tend to have neighbours with more sides, and vice-versa is captured by the so-called Aboav-Weaire's law \citep{aboav1970,aboav1980,weaire1974}
\begin{equation}
\mu_{m(n)}=6-\gamma+(6\gamma+\sigma^2_n)/n.
\end{equation}
It describes the correlation between $n$, and the average number of neighbors of cells $m(n)$ adjacent to ones with $n$ neighbors. Here, $\sigma^2_n$ is the second central moment of the distribution of $n$, and $\gamma$ is a constant that may decrease as $\sigma^2_n$ increases \cite{caer1993,vincze2004}, or may be independent of $\mu_2$ \cite{saraiva2009}.

Current analysis shows  a relatively large range of validity of the Aboav-Weaire's law, with $\gamma$ being independent of $\sigma^2_n$ (which decreases with $\phi_g$), but sensitive to the elongation of the packed ellipses (Fig. 8d). Deviations from  Aboav-Weaire's law become more important for strongly elongated packed objects at high packing fractions (circular symbols in the top panel of Fig. 8d). Furthermore, for discoid shapes, strong departures from linearity are observed for $n>8$ (bottom panel of Fig. 8d). It is at this stage not clear if this observation is a result of sample size or a true nature of the assembly. 

\section{Conclusions}

In this manuscript, we perform a systematic analysis of assemblies of ellipses generated in a stochastic process, over the entire parameter range of packing fractions and elongations, including both fluid (non-static) states and jammed (mechanically static) configurations. We focus on the analysis of measures that have been in the past used in experiments, and besides their distributions, we analyse the correlations between various measures. Given the fact that this analysis is focused only on 2D assemblies, in which the generating protocol does not play a major role, the obtained results should posses a certain level of universality. This is important not only because the geometric nature of the assemblies is fully captured, but also because the obtained trends may be particularly useful in the analysis of experimental data in 2D. Namely, strong correlations can be used as a criterion to focus on a particular measure, which is easily accessible, knowing that its distribution stores the same information as some other. 

In the performed analysis, we were able to link characteristic behaviour of the distributions and correlations to particular features of the assemblies. Several well known results were recovered, validating our calculations. Consistently with previous reports, we find that at low packing fraction, shape of the packed particles does not influence the structure of the assemblies while at the high packing fractions morphology of the Voronoi cells depends highly on the shape of the packed ellipses. 

We can observe another interesting phenomenon that is  common to all morphological measures. Specifically, all the standardized distributions at all packing fractions overlap except for circle-like particles at high packing fractions. The explanation for the previous event is that packing fraction of $0.8$  is in random close assembly of circles range ($\phi_g = 0.82 \pm 0.02$ \citep{berryman1983}) and therefore, substantial regularities in our systems start to show when the packing fraction is at 0.8 or above.

We hope this work will provide a framework for the analysis of  assemblies with more complex interactions, which are now starting to be investigated. This does not refer only to assemblies of soft objects, but also active particles \citep{sharp2019,yang2017,merkel2018,merkel2019,dasgupta2018}. In the latter case it is very important to delineate the geometric effects of density and shape, which have been meticulously studied herein, from the effects of activity.  By finding deviations from the  results presented in this manuscript, one could clearly identify such features of the assemblies.     
\section*{Conflicts of interest}

There are no conflicts to declare.

\section*{Acknowledgments}

We thank Tomaso Aste for useful discussions. ASS and GST thank the German Academic Exchange Service and Universities Australia for travel funding through a collaborative grant scheme. JL, SK and ASS acknowledge the support of the European Research Council (ERC) under grant ERC StGMembranesAct 2013-33728 and the support from the Excellence Cluster: Engineering of Advanced Materials at the FAU Erlangen-N\"urnberg.

\bibliography{2Dpacking} 
\bibliographystyle{rsc} 

\end{document}